\documentclass[manuscript]{acmart}
\AtBeginDocument{%
  }

\setcopyright{rightsretained}
\copyrightyear{2025}
\acmYear{2025}
\acmDOI{XXXXXXX.XXXXXXX}
\acmISBN{978-1-4503-XXXX-X/2018/06}

\usepackage{algorithm}
\usepackage{algorithmic}
\usepackage{amsmath}
\usepackage{subfigure}
\usepackage{xcolor}
\usepackage{colortbl}
\usepackage{xspace}

\begin{document}

\title{Learning in Context: Personalizing Educational Content with Large Language Models to Enhance Student Learning}


\author{Joy Jia Yin Lim}
\orcid{0009-0006-7971-6096}
\author{Daniel Zhang-Li}
\orcid{0009-0009-3681-1896}
\author{Jifan Yu}
\orcid{0000-0003-3430-4048}
\author{Xin Cong}
\author{Ye He}
\author{Zhiyuan Liu}
\author{Huiqin Liu}
\author{Lei Hou}
\author{Juanzi Li}
\author{Bin Xu}
\authornotemark[2]
\email{lin-jy23@mails.tsinghua.edu.cn}
\email{zlnn23@mails.tsinghua.edu.cn}
\email{yujifan@tsinghua.edu.cn}
\email{congxin1995@tsinghua.edu.cn}
\email{heye23@mails.tsinghua.edu.cn}
\email{liuzy@tsinghua.edu.cn}
\email{liuhq@tsinghua.edu.cn}
\email{houlei@tsinghua.edu.cn}
\email{lijuanzi@tsinghua.edu.cn}
\email{xubin@tsinghua.edu.cn}
\affiliation{
  \institution{Tsinghua University}
  \city{Beijing}
  \country{China}
}

\renewcommand{\shortauthors}{Joy Jia Yin Lim et al.}
\newcommand{\system}{{\textit{PAGE}}\xspace}

\begin{abstract}
   Standardized, one-size-fits-all educational content often fails to connect with students' individual backgrounds and interests, leading to disengagement and a perceived lack of relevance. To address this challenge, we introduce \system, a novel framework that leverages large language models (LLMs) to automatically personalize educational materials by adapting them to each student's unique context, such as their major and personal interests. To validate our approach, we deployed \system in a semester-long intelligent tutoring system and conducted a user study to evaluate its impact in an authentic educational setting. Our findings show that students who received personalized content demonstrated significantly improved learning outcomes and reported higher levels of engagement, perceived relevance, and trust compared to those who used standardized materials. This work demonstrates the practical value of LLM-powered personalization and offers key design implications for creating more effective, engaging, and trustworthy educational experiences.
\end{abstract}




\maketitle

\section{Introduction}

\begin{figure}[ht]
  \centering
  \includegraphics[width=0.9\linewidth]{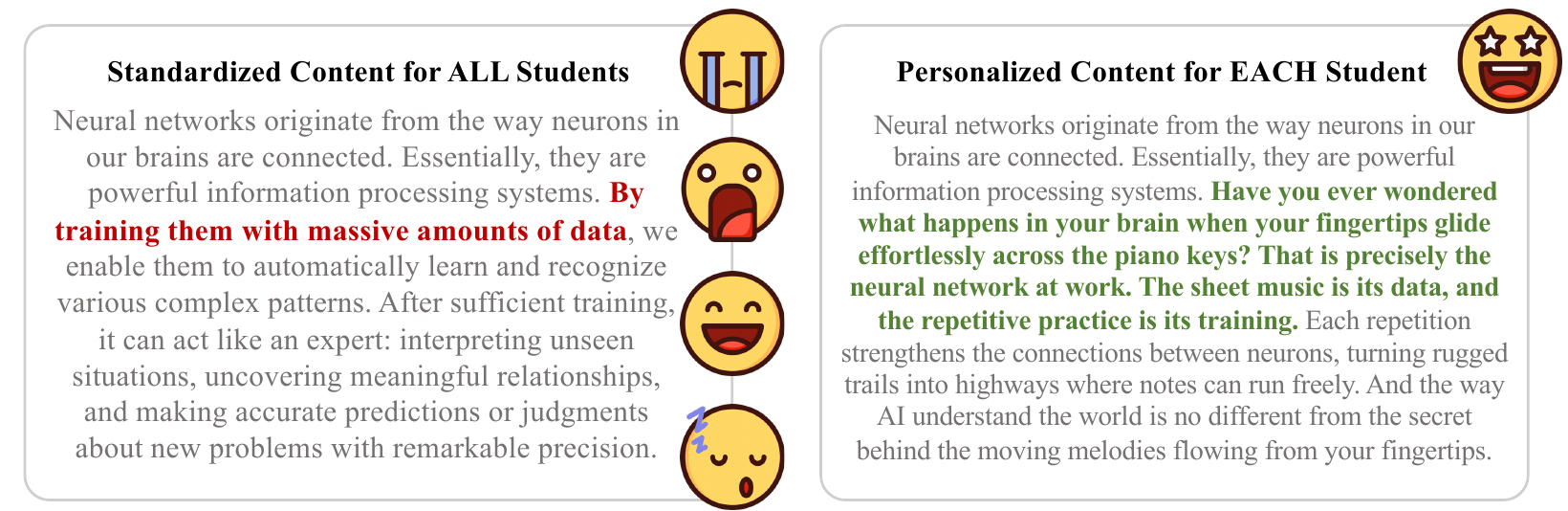}
  \caption{Comparison between standardized content for all students in traditional educational paradigms (including physical classrooms and online learning platforms) and personalized content for individual students.}
  \label{fig:intro-case}
\end{figure}

Standardized, one-size-fits-all educational content is failing to engage students~\cite{wu2024comprehensiveexplorationpersonalizedlearning}. As shown in Figure~\ref{fig:intro-case}, in both traditional classrooms and dominant online learning platforms, learners are often presented with generic materials that disregard their individual backgrounds, interests, and real-world contexts~\cite{collins2010second,collins2018rethinking}. This disconnect frequently leads to a "crisis of relevance" and "affective disengagement," where students struggle to see the value of abstract knowledge in their own lives~\cite{Yang_2019,bondie2019does}. While the vision of personalized education has been a long-standing goal in educational technology~\cite{wu2024comprehensiveexplorationpersonalizedlearning,maghsudi2021personalized,taber2008conceptual}, the resource-intensive nature of preparing personalized content have been a major barrier to its widespread adoption~\cite{bernacki2018role}. This leaves a critical gap between the potential of personalized learning and the reality of current educational practices, resulting in diminished student motivation and learning outcomes~\cite{smith2004instructional,pekrun2006control,pekrun2014control}.

Recent advances in large language models (LLMs) present new possibilities to deliver personalized learning at scale~\cite{walkington2014motivating,harackiewicz2016interest,reber2018personalized,bernacki2021systematic}, with their remarkable capabilities in contextual understanding, content generation, and knowledge synthesis~\cite{pelaez2024impact,gan2023large}.
While numerous studies have explored the potential of LLMs in various educational applications~\cite{waldeck2007answering,ramesh2024llm_personalization}, existing approaches often provide only superficial personalization, such as adjusting content complexity, rather than achieving deep contextual adaptation~\cite{he2024evaluating,chen2023dynamicprofile,park2024empowering}. This technological gap points to a significant design challenge for the community: \emph{How can we leverage the power of LLMs to create educational experiences that are not just algorithmically optimized, but are also genuinely resonant and meaningful to each individual?}

To address this challenge, we introduce \system, a retrieval-augmented framework for \textbf{P}ersonalized educ\textbf{A}tional content \textbf{GE}neration. Our research aims to explore the potential of LLMs in generating educational content with high ecological validity and pedagogical efficacy, by deep understanding and adapting to each student's unique context. 
Building upon the demonstrated effectiveness of high-quality, human-authored educational materials, we investigate whether and how LLM-generated personalized content can further enhance student learning experience. We specifically focus on the following research questions:

\begin{itemize}
    \item \textbf{RQ1}: To what extent does interacting with context-aware personalized content affect students' knowledge acquisition compared to standardized materials?
    \item \textbf{RQ2}: How does the personalized content shape a student's engagement, their perception of the material's relevance, and their trust in the learning system?
    \item \textbf{RQ3}: What are the mechanisms through which this personalization influences a student's learning process and their overall perception of the educational experience?
\end{itemize}

To answer these questions, we conducted a comprehensive user study with $40$ university students from diverse academic backgrounds. We integrated \system into a semester-long intelligent tutoring system~\cite{yu2024mooc} and compared the effects of the personalized content generated by our system against human-authored standardized materials. We gathered both quantitative data on learning outcomes and rich qualitative insights into the students' subjective experiences.

Our contribution are as follows:
\begin{enumerate}
    \item We introduce \system, a novel framework that generates personalized educational content by adapting standardized materials to a student's unique background and interests.
    \item We conduct comprehensive user study and provide robust evidence that our personalized content improves both learning outcomes and student engagement when compared to high-quality standardized materials.
    \item We demonstrate the practical applicability of LLM-generated content in an authentic, real-world educational setting, distilling key design implications for creating trustworthy, context-aware learning experience.
\end{enumerate}

\section{Related Work}

This section briefly reviews the theoretical foundations of personalized content generation, the impact of LLMs for enhanced personalization, and the development of intelligent tutoring systems. 

\subsection{Personalized Content Generation}
Personalization aims to tailor educational content to the unique needs, goals, and contexts of individual students~\cite{king2017reimagining,alamri2020using,belenky2014effects}. Early attempts explored a variety of applications, including template-based keyword insertion~\cite{hogheim2015supporting,hogheim2017eliciting}, character customization in educational media~\cite{pataranutaporn2022ai}, and static or rule-based student modeling~\cite{ku2007effects,lopez1992effect}. While these methods demonstrated potential applicability, they often 
required considerable manual effort, lacked scalability, and struggled to capture the complexity and variability of student contexts~\cite{frenzel2012beyond}. We aim to address these limitations by advancing scalable, deeper personalization to enable more adaptive learning experiences.

\subsection{LLMs for Enhancing Personalization}
The recent advent of LLMs presents transformative potential for personalized education~\cite{wen2024ai,razafinirina2024pedagogical}.
Some reframed standard problems (e.g., algebra word problems) using contexts derived from student interests~\cite{yadav2023contextualizing}, while others focused on improving specific tasks such as vocabulary learning~\cite{leong2024putting}. Systems like PACE~\cite{liu2025one} leveraged LLMs to create personalized conversational tutors for predefined learning styles. Despite the promising results, existing approaches have focused on domain-specific applications~\cite{leong2024putting} or particular aspects of personalization, such as surface-level contextualization or adaptation using predefined learning style models. Less attention has been paid to dynamically adapting standardized content to individual learning contexts. Our work aims to address this gap.

\subsection{Intelligent Tutoring Systems (ITS)}
ITS represents a significant advancement in personalized education~\cite{nye2014autotutor,bubeck2023sparksartificialgeneralintelligence}, as it created interactive learning environments that emulate one-on-one human tutoring~\cite{nye2014autotutor}. Existing ITS often implement personalization through: dynamically adjusting task sequence and difficulty~\cite{park2024empowering,chen2024empowering,zhang2024awaking,zhang2024simulating}; providing contextually tailored feedback~\cite{pal2024autotutor,zhou2025study,golchin2024large}; and optimizing learning trajectories through structured curricula~\cite{goslen2024llm,park2023generative,manoharan2021maximizing,hu2024teaching}. While effective in constrained domains, they relied on domain-specific or pre-authored content~\cite{pardos2024chatgpt,murray2003overview}. The development of a comprehensive, course-wide framework that can adapt to heterogeneous domains and diverse learning requirements remains largely underexplored.
\definecolor{lightpink}{RGB}{249, 235, 235}
\definecolor{lightblue}{RGB}{232, 245, 252}
\definecolor{peachpink}{RGB}{249, 235, 235}

\section{Personalized Educational Content Generation (\system)}

\begin{figure*}[t]
  \centering
  \includegraphics[width=0.95\linewidth]{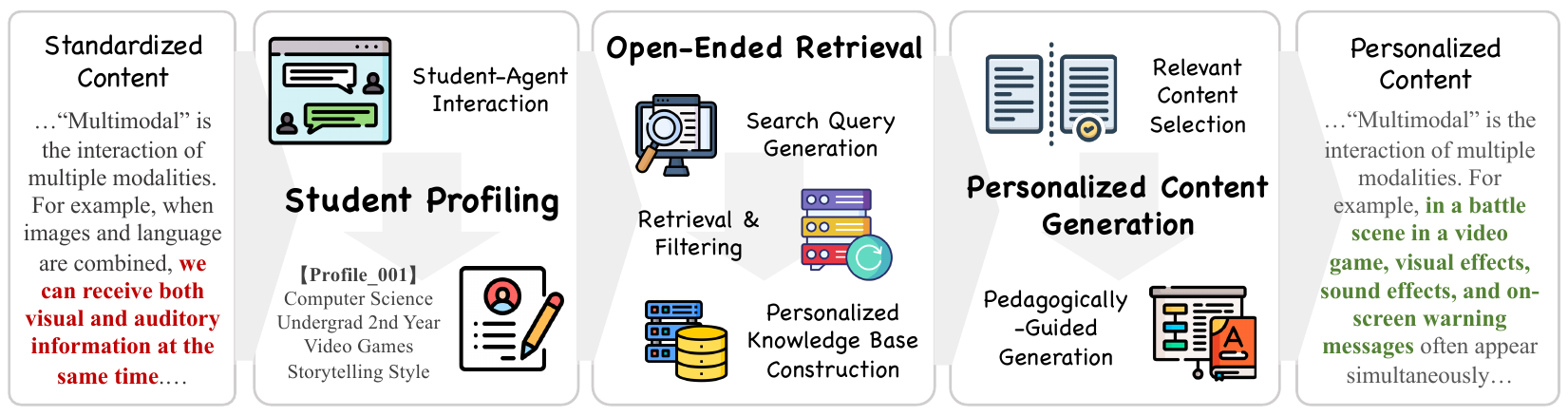}
  \caption{\system incorporates: (1) Cognitive-Affective Student Profiling, which captures cognitive attributes and affective characteristics; (2) Open-Ended Knowledge Retrieval, which retrieves relevant knowledge aligned with individual contexts; and (3) Pedagogically-Guided Adaptation, which generates personalized content guided by established pedagogical principles.}
  \label{fig:main}
\end{figure*}

This section details the methodological architecture of \system, as shown in figure~\ref{fig:main}, including (1) cognitive-affective student profiling, (2) adaptive knowledge retrieval, and (3) pedagogically-guided content adaptation.


\subsection{Cognitive-Affective Student Profiling}
\label{sec:student-profiling}
A well-structured student profile serves as the foundation for any effective personalized learning system~\cite{luis2013}. Inspired by the principles of \textbf{Universal Design for Learning (UDL)}~\cite{rose2000universal}, an established theory for creating adaptive learning environments, our profilling approach is designed to proactively support student diversity, by identifying and accommodating variability in how students engage with content, process information, and demonstrate understanding. To implement this, we employ a conversational agent that engages in dialogues with students before and after each class session (Appendix~\ref{sec:system_prompts}). This interaction allows our system to dynamically construct and update a profile integrating both \textbf{cognitive} (e.g., academic major, year of study) and \textbf{affective} features (e.g., interests, learning preferences). Cognitive features from institutional data determine a student's knowledge level, while affective features from self-reports gauge their motivation and engagement. By embedding both cognitive and affective features, our profiling approach enables the generation of educational content that is not only personalized, but also inclusive, responsive, and pedagogically effective. Table~\ref{tab:student_profile_example} presents an anonymized example of a student profile constructed by our system.

\subsection{Open-Ended Knowledge Retrieval}
To create a flexible knowledge structure that adapts to diverse contexts without relying solely on LLM parametric knowledge or static knowledge bases, we implement an open-ended retrieval mechanism that identifies and retrieves materials relevant to each student profile. 

\subsubsection{Profile-Aware Query Generation}
To begin with, we leverage an LLM (Appendix~\ref{sec:system_prompts}) to analyze the student profile (denoted as $P$) and the target content (denoted as $s_i$), generating search queries that strategically integrate profile attributes with specific content topics. For example, when explaining the concept of "Multimodality" to a computer science student interested in video games, the system may generate contextual queries such as "\emph{application of AI modalities in game development}" or "\emph{multimodal interaction in video gaming}."

\subsubsection{Retrieval and Filtering}
The generated queries are executed sequentially via search engines (e.g., Google~\footnote{https://www.google.com}, Bing~\footnote{https://www.bing.com}), with results filtered by a prioritization algorithm that favors educational and scholarly sources. All documents are processed to remove advertisements and irrelevant materials, preserving only substantive information.

\subsubsection{Personalized Knowledge Base Construction}
The retrieved documents (denoted as $D_{s_i,P}$) are segmented into semantically coherent knowledge chunks (denoted as $T_{s_i,P}$). These chunks are transformed using an embedding function $\phi(\cdot)$ and stored in a vector database, creating a personalized knowledge base $K_{s_i,P} = \{\phi(t_{ij}) \mid t_{ij} \in T_{s_i,P} \}$ for each student. This individualized repository serves as the foundation for subsequent content adaptation, enabling precise retrieval of contextually relevant material during personalization.

\subsubsection{Similarity-based Knowledge Selection}
For each standard content segment $s_i$, we identify the most relevant information by calculating semantic similarity between the content embedding $\phi(s_i)$ and the embeddings of knowledge chunks $\phi(t_{ij})$. Using cosine similarity as our metric, we select the top-$k$ most relevant chunks (empirically set $k=5$) to form the retrieval set (denoted as $K_{S,P}^{(k)}(s_i)$), which serves as supplementary context for the generation process.

\begin{table}[t]
  \centering
  \small 
  \caption{An example of a student profile constructed by an LLM-powered personal agent. The agent captures both structured academic data and unstructured interests from dialogue, which it then processes into structured tags.}
  \label{tab:student_profile_example}
  \begin{tabular}{p{0.9\columnwidth}} 
    \toprule
    \rowcolor{lightpink} \textbf{\textit{1. Basic Information}} \\
    \addlinespace 
    \textbf{Student ID:} student\_001 \\
    \textbf{Profile Updated:} 2025-08-24 \\ 
    \addlinespace 
    \midrule
    \rowcolor{lightblue} \textbf{\textit{2. Academic Profile}} \\
    \addlinespace 
    \textbf{Year:} Sophomore \\
    \textbf{Major:} Computer Science and Technology \\
    \addlinespace 
    \midrule
    \rowcolor{peachpink} \textbf{\textit{3. Interest Profile}} \\
    \addlinespace 
    \textbf{Raw Text Input (Captured from Dialogue):} \\
    "I like to play games, mainly single-player RPGs with good plots. I'm currently playing 'Baldur's Gate 3,' and I feel the narrative and world-building are amazing. Nothing else to add." \\
    \addlinespace 
    \textbf{Structured Tags (Inferred by LLM):} \\
    \quad \textbf{Domain:} Entertainment \\
    \quad \textbf{Category:} Gaming \\
    \quad \textbf{Sub-Category:} Single-Player RPG \\
    \quad \textbf{Keywords:} Baldur's Gate 3, Story Narrative, World-Building, Role-Playing \\
    \addlinespace 
    \bottomrule
  \end{tabular}
\end{table}

\subsection{Pedagogically-Guided Content Adaptation.}
The personalization of educational content employs a retrieval augmented generation (RAG) process that integrates contextually relevant information from the curated knowledge base. 
To ensure pedagogical soundness, we systematically design the content adaptation framework grounded in established pedagogical principles, aiming to balance the flexibility of LLM-generated content with the rigor of evidence-based theories.

\subsubsection{Content Selection and Filtering.}
We first apply rule-based checks to exclude contents that are overly brief, elementary, introductory, or concluding in nature, preserving these elements in their original standard form without personalization. This selective approach prevents the generation of redundant or pedagogically inappropriate content that could hinder student engagement and dilute the overall instructional value of the material.

\subsubsection{Theoretical Framework for Content Adaptation}
Our content adaptation strategy draws upon three foundational pedagogical theories to guide the personalization process, as operationalized in the prompt template shown in Table~\ref{tab:generation_prompt}. 

\paragraph{\textbf{Bloom's Taxonomy}} Drawing on Bloom's Taxonomy~\cite{bloom1956taxonomy}, we map standardized content across the spectrum of cognitive levels, from basic recall and comprehension to higher-order thinking skills such as analysis and synthesis, enabling a precise alignment between content complexity and individual readiness.

\paragraph{\textbf{Vygotsky's Zone of Proximal Development (ZPD)}} We apply ZPD~\cite{vygotsky1978mind} theory to strategically identify optimal content segments for personalization. We focus adaptation efforts on those that challenge students just beyond their current mastery level while remaining accessible with appropriate scaffolding. 

\paragraph{\textbf{Universal Design for Learning (UDL)}} We further incorporate UDL~\cite{rose2000universal} principles to create multidimensional adaptations that serve diverse needs, spanning dimensions such as adjustments to conceptual depth, provision of varied illustrative examples, and development of contextually relevant explanatory analogies.  

\begin{table*}[t]
  \centering
  \small
  \caption{The structured prompt guiding the LLM for personalized content generation. The prompt is framed with foundational pedagogical theories and a strict three-step chain-of-thought process to ensure high-quality, instructionally sound output.}
  \label{tab:generation_prompt}
    \begin{tabular}{p{0.9\textwidth}}
    \toprule
    \textbf{Prompt Template for Personalized Content Generation} \\
    \midrule
    \addlinespace
    You are an pedagogical expert specializing in educational content personalization. Your task is to adapt a standardized lecture content for a specific student based on their profile, guided by established pedagogical principles. \\
    \addlinespace
    \textbf{\#\# INPUTS:} \\
    \texttt{\{student\_profile\}} \\
    \texttt{\{retrieved\_documents\}} \\
    \texttt{<script>\{standardized\_content\}</script>} \\
    \addlinespace
    \textbf{\#\# INSTRUCTIONS:} \\
    Your content adaptation strategy must be guided by the following three foundational pedagogical theories:\\
    $\bullet$ \emph{Bloom's Taxonomy:} \texttt{\{Detail Instructions\}}\\
    $\bullet$ \emph{Vygotsky's Zone of Proximal Development (ZPD):} \texttt{\{Detail Instructions\}}\\
    $\bullet$ \emph{Universal Design for Learning (UDL):} \texttt{\{Detail Instructions\}} \\
    \addlinespace
    \textbf{Strictly follow this three-step chain-of-thought process:} \\
    \textbf{Step 1: Assess the Need for Personalization} \\
    $1.$ Analyze ONLY the content within the \texttt{<script>} tags.\\ 
    $2.$ If the content is very short (1-2 sentences), purely transitional, or contains no complex concepts, you must immediately halt and output only the string \texttt{[None]}. \\
    \addlinespace
    \textbf{Step 2: Execute Personalization (If applicable)} \\
    $1.$ If adaptation is needed, introduce a maximum of 0-2 subtle adjustments.\\
    $2.$ \textbf{DO:} Add relevant background, analogies, or examples connected to the student's profile. \\
    $3.$ \textbf{DO NOT:} Overwrite or delete core concepts, alter the logical structure, or add generic "filler" content.\\ 
    \addlinespace
    \textbf{Step 3: Review for Coherence and Neutrality}\\
    $1.$ Ensure the final script is logically coherent and reads naturally.\\
    $2.$ \textbf{CRITICAL:} Remove any explicit traces of personalization (e.g., no phrases like "Based on your interest in...").\\
    \addlinespace
    \textbf{\#\# FINAL OUTPUT REQUIREMENTS:} \\
    \texttt{\{Detailed Output Requirements\}}\\
    \addlinespace
    \bottomrule
  \end{tabular}
\end{table*}

\section{Evaluation}
\label{sec:experiments}


In this section, we conducted a two-phase evaluation to understand the impact of \system. We first performed an expert-based evaluation to validate the pedagogical quality of the generated content (Study 1). After confirming its quality, we conducted an in-situ user study with university students ($n=40$) to investigate the effects of personalization on their learning outcomes and subjective experiences (Study 2).


\definecolor{palepink}{RGB}{180, 100, 100}   
\definecolor{lightpink}{RGB}{190, 90, 120}  
\definecolor{lightblue}{RGB}{80, 130, 170}  
\definecolor{palpeach}{RGB}{190, 140, 120}   
\definecolor{lightgrey}{RGB}{100, 100, 100}  
\definecolor{lightpalepink}{RGB}{245, 230, 230}
\definecolor{lightpalpeach}{RGB}{250, 240, 235}
\definecolor{lighterpink}{RGB}{250, 235, 240}
\definecolor{lighterlightgrey}{RGB}{240, 240, 240}
\definecolor{lighterlightblue}{RGB}{235, 245, 250}

\subsection{Study 1: Expert Evaluation of Pedagogical Quality}

Before the deployment with students, we need to address a critical question: \emph{Is the generated content pedagogically valid?} To answer this, we conducted an evaluation with domain experts.

\subsubsection{Materials}
To create a challenging and representative testbed, we selected instructional materials from five diverse university-level courses: \emph{Towards Artificial General Intelligence (TAGI, Computer Science)}, \emph{How to Study in University (HSU, Education)}, \emph{Biology (BIO, Natural Science)}, \emph{Probability Theory and Mathematical Statistics (PTMS, Mathematics)}, and \emph{Psychology (PSY, Psychology)}. For each course, we selected core, self-contained sections from validated, human-authored curricula. We then generated a set of unique student profiles for each section, incorporating a wide range of academic backgrounds and personal interests. All profiles were systematically created based on representative university enrollment data and then manually validated by our research team to ensure realism. See Table~\ref{tab:dataset} for detailed statistics.

\begin{table}[t]
  \centering
  \caption{Dataset Statistics, including sample counts, word counts, query counts, and retrieved document counts.}
  \begin{tabular}{c|cccccc}
    \toprule
    Course & \# Samples & \# Words & \# Queries & \# Retri. Docs \\
    \midrule
    TAGI   & 20  & 8138  & 168 & 699  \\
    HSU    & 10  & 2906  & 80  & 416  \\
    BIO    & 10  & 2189  & 84  & 499  \\
    PTMS   & 10  & 3007  & 73  & 460  \\
    PSY    & 10  & 1566  & 84  & 499  \\
    \midrule
    Total  & 60  & 17806 & 489 & 2573   \\
    \bottomrule
    \end{tabular}
    \label{tab:dataset}
\end{table}

\subsubsection{Experimental Conditions}
To rigorously evaluate the content generated by \system, we compared it against a human-authored baseline and three alternative LLM generation methods. This allowed us to assess not only the overall effectiveness of our system, but also to isolate the contributions of its core components (i.e., the backbone model and the retrieval strategy). Specifically, we evaluated content from the following five conditions:

\begin{enumerate}
    \item \textcolor{lightblue}{\textbf{\system}}: Personalized content generated by our proposed retrieval-augmented framework.
    \item \textcolor{palepink}{\textbf{Human}}: The original content from the course materials, serving as our control group and quality baseline.
    \item \textcolor{palpeach}{\textbf{4o (w/ RAG)}}: Personalized content generated by \emph{GPT-4o}~\cite{hurst2024gpt} , with same retrieval-augmentated strategy.
    \item \textcolor{lightpink}{\textbf{o1 (w/ RAG)}}: Personalized content generated similarly by \emph{OpenAI-o1}~\cite{jaech2024openai} model.
    \item \textcolor{lightgrey}{\textbf{r1 (w/o RAG)}}: An ablation condition, using the same backbone model as \system(\emph{DeepSeek-r1}~\cite{guo2025deepseek}) but without the retrieval mechanism. This allows us to directly measure the impact of retrieval component.
\end{enumerate}

\subsubsection{Evaluation Metrics}
To structure the expert evaluation, we developed a rubric based on six key dimensions. We adapted four established metrics from prior work on multimedia instructional design~\cite{leceval2025} and introduced two new metrics specifically to assess the quality of LLM-driven personalization:
\begin{enumerate}
    \item \textbf{Instructional Accuracy:} Assesses the factual correctness and completeness of the core concepts.
    \item \textbf{Expressive Clarity:} Evaluates whether the language is clear, concise, and easy to understand.
    \item \textbf{Logical Coherence:} Measures the logical flow and consistency of the presented arguments and ideas.
    \item \textbf{Student Engagement:} Rates the content's potential to capture and maintain a student's interest and motivation.
    \item \textbf{Linguistic Naturalness:} Assesses whether the style and tone sound like natural, human-written language.
    \item \textbf{Personalization Relevance:} Measures how effectively and appropriately the content is tailored to the student's specific profile  (e.g., academic backgrounds and personal interests).
\end{enumerate}

\subsubsection{Procedure}
We recruited five experts with backgrounds in pedagogical research and instructional design to rank the content from different conditions on our predefined metrics, and provide written justification for their rankings. Detailed guidelines are shown in Appendix~\ref{sec:annotation}. To mitigate bias, the evaluation was conducted as a blind review, where the experts were unaware of the source of the contents. Each content was independently reviewed and ranked by two different experts. This process yielded a high inter-annotator agreement (Kendall's $\alpha \geq 0.8$), confirming the reliability and consistency. The final rankings were then converted into numerical scores for quantitative analysis.

\subsubsection{Key Findings}

Our expert evaluation revealed three key findings regarding the quality of LLM-generated content.

\textbf{1. Personalized Content Outperforms Standardized Materials in Engagement and Relevance.}
Table~\ref{tab:offline} showed a substantial gap between personalized content from \system and the standardized baseline. Experts consistently ranked \system as significantly better in \emph{Personalization Relevance} ($92.2$ vs. $33.8$) and \emph{Student Engagement} ($87.0$ vs. $35.9$). Interestingly, experts also preferred \system's \emph{Linguistic Naturalness} over the real human-authored text ($86.0$ vs. $33.8$). According to the qualitative feedback, this was not because the LLM was more "human", but because standardized academic writing was often perceived as "dry" and "impersonal". By weaving in relevant analogies, \system produced a communication style that was judged to be more natural and accessible for learners.

\textbf{2. Retrieval-Augmentation is Crucial for Factual Trustworthiness and Clarity.} Our ablation study, which compared \system to a version without its retrieval component, demonstrated that grounding the LLM in external knowledge is essential. Without retrieval, scores for \emph{Instructional Accuracy} dropped sharply from $77.8$ to $53.7$ and \emph{Expressive Clarity} from $75.5$ to $53.0$. From an HCI perspective, this finding underscores that retrieval-augmentation is a critical mechanism for ensuring the trustworthiness of the generated content, by mitigating factual errors and hallucinations. Furthermore, by providing richer and more reliable information, LLM can generate clearer examples and explanations, which directly helps to reduce student cognitive load, especially when tackling complex topics.

\textbf{3. Foundation Model Choice is a Key Design Decision That Shapes User Experience.} Our comparison of different foundation models showed that the choice of backbone LLM is not only a technical detail but a significant design decision. While all retrieval-augmented models performed well, the \emph{DeepSeek-r1} model used in \system consistently outperformed the others on the more affective, user-facing dimensions: \emph{Student Engagement}, \emph{Linguistic Naturalness}, and \emph{Personalization Relevance}. This suggests that for educational systems, foundation model is a crucial design choice that shapes the relational and qualitative aspects of the learning experience.

\begin{table}[t]
  \centering
  \caption{Results for the benchmark evaluation, showing the average scores across six evaluation dimensions.}
  \begin{tabular}{c|cccccc|c}
    \toprule
    Method & Instr. & Expre. & Coher. & Engag. & Natur. & Perso. & Overall \\
    \midrule
    \rowcolor{lightpalepink} \emph{Human-Authored} & $55.5$ & $56.5$ & \underline{$57.9$} & $35.9$ & $33.8$ & $33.8$ & $45.5$ \\
    \midrule
    \rowcolor{lightpalpeach} \emph{4o (w/ RAG)} & $55.9$ & $56.8$ & $57.2$ & $55.8$ & $54.1$ & $54.8$ & $55.9$ \\
    \rowcolor{lighterpink} \emph{o1 (w/ RAG)} & \underline{$57.1$} & \underline{$58.2$} & $55.2$ & $54.0$ & $57.1$ & $53.4$ & $55.8$ \\
    \rowcolor{lighterlightgrey} \emph{r1 (w/o RAG)} & $53.7$ & $53.0$ & $53.4$ & \underline{$67.3$} & \underline{$69.0$} & \underline{$65.7$} & \underline{$60.3$} \\
    \midrule
    \rowcolor{lighterlightblue} \emph{\system} & \textbf{77.8} & \textbf{75.5} & \textbf{76.2} & \textbf{87.0} & \textbf{86.0} & \textbf{92.2} & \textbf{82.4} \\
    \bottomrule
  \end{tabular}
  \label{tab:offline}
\end{table}

\definecolor{lightpink}{RGB}{249, 235, 235}
\definecolor{lightblue}{RGB}{232, 245, 252}
\definecolor{peachpink}{RGB}{249, 235, 235}

\subsection{Study 2:  In-Situ Evaluation of the Student Learning Experience}

Building on the validated pedagogical quality of our personalized content in Study 1, we aimed to understand its real-world impact on the student learning experience. To achieve this, we conducted an in-situ user study, deploying \system within a live intelligent tutoring system used by university students. Our investigation was guided by three research questions that examine the system's impact from different perspective:
\begin{itemize}
    \item \textbf{RQ1}: To what extent does interacting with context-aware personalized content affect students' knowledge acquisition compared to standardized materials?
    \item \textbf{RQ2}: How does the personalized content shape a student's engagement, their perception of the material's relevance, and their trust in the learning system?
    \item \textbf{RQ3}: What are the mechanisms through which this personalization influences a student's learning process and their overall perception of the educational experience?
\end{itemize}

\subsubsection{Study Design}

To investigate our research questions, we employed a between-subjects experimental design. Participands were randomly assigned to one of two conditions based on \emph{Content Type}: 
\begin{enumerate}
    \item \textbf{Personalized:} Personalized educational content generated in real-time by \system.
    \item \textbf{Standardized:} The validated, human-authored educational material from the original courses.
\end{enumerate}

\paragraph{Materials} Our study was conducted within two well-established introductory university courses: \emph{Towards Artificial General Intelligence (TAGI)} and \emph{How to Study in University (HSU)}, for their validated, high-quality design and broad accessibility (requiring no specialized prior knowledge). This provided a robust learning environment that allowed us to isolate the effects of personalization across a diverse student population.

\begin{figure}[ht]
  \centering
  \includegraphics[width=0.7\linewidth]{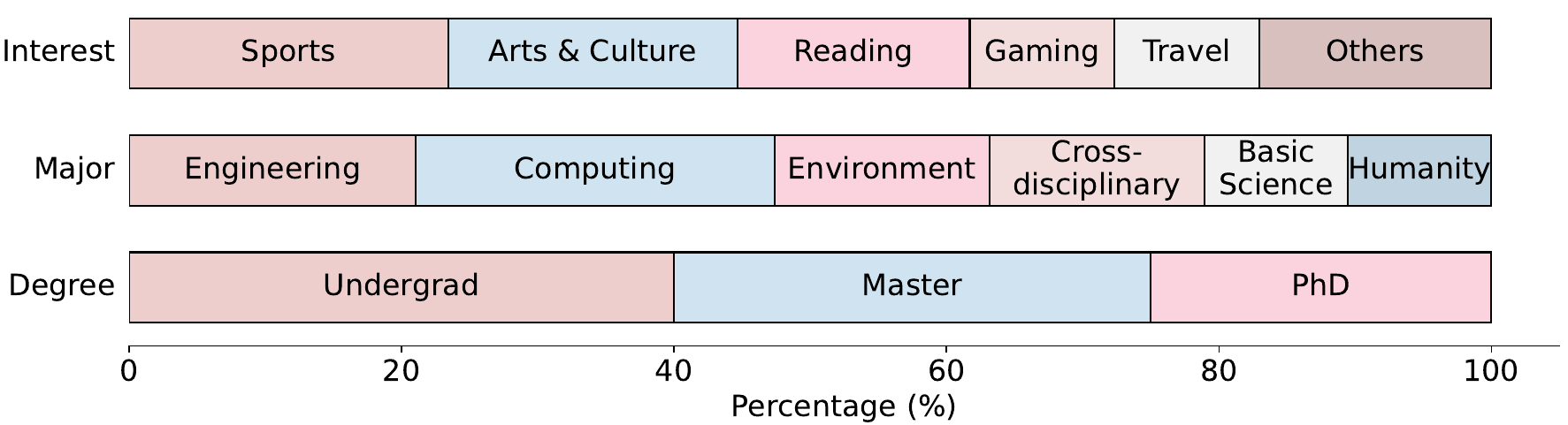}
  \caption{Distribution of participant demographics, illustrating diversity in academic backgrounds and interests.}
  \label{fig:pie-statistics}
\end{figure}

\paragraph{Participants} We recruited 40 university students ($N=40$) from a wide range of academic backgrounds (Figure~\ref{fig:pie-statistics}). Participants were randomly assigned to either the \emph{Personalized} condition ($n=20$) or the \emph{Standardized} condition ($n=20$). The assignment was balanced across both courses, with $10$ students per condition in each course.

\paragraph{Measures} We employed a mixed-methods approach to develop a comprehensive understanding of student experience, combining quantitative metrics with qualitative insights from semi-structured interviews. 

\textbf{Quantitative Measures.} To objectively measure the impact of personalized, we collected the following data:
\begin{itemize}
    \item \textbf{Knowledge Assessment (RQ1)}: Participants completed a 15-minute, instructor-designed knowledge test after their learning session to measure learning outcomes. The test included multi-choice questions to evaluate concept recognition and short-answer questions to assess knowledge application.
    \item \textbf{Post-Session Questionnaire (RQ2)}: Participants rated their subjective experience on a post-session questionnaire using 7-point Likert scales. The questionnaire included four scales adapted from the widely-used User Experience Questionnaire (UEQ) to measure user experience: (1) \emph{Attractiveness} (overall appeal); (2) \emph{Stimulation} (engagement and motivation); (3) \emph{Efficiency} (ease and speed of understanding); and (4) \emph{Dependability} (confidence in accuracy and reliability), and two independent scale derived from prior work~\cite{zhang2024awaking} to assess Perceived Learning: (5) \emph{Learning New Concepts} (perceived understanding of new concepts) and (6) \emph{Deepening Understanding} (perceived depth of understanding of complex concepts). The complete questionnaire is available in Table~\ref{tab:questionnaire}.
\end{itemize}

\textbf{Qualitative Measures.} To understand the why behind our quantitative results (RQ3), we conducted a semi-structured interview with each participant immediately after the session. Key questions included: \textit{"Can you recall a moment where the material felt particularly clear or connected to your interests?"} and \textit{"How did this experience compare to how you typically learn new subjects?"}

\begin{table}[t]
  \small
  \centering
  \caption{The complete survey questionnaire, including three distinct sections: (1) Learning Outcome; (2) Learning Experience; (3) Open-Ended Question.}
  \begin{tabular}{p{0.95\columnwidth}}
    \toprule
    \rowcolor{lightpink} \emph{Section 1: Learning Outcome} \\
    1. The explanations were simple and easy to understand. \\
    2. The platform helped me gain a deeper understanding of the content. \\
    3. \{ Knowledge Assessment \} \\
    \midrule
    \rowcolor{lightblue} \emph{Section 2: Learning Experience} \\
    4. The learning process on the platform was engaging and interesting. \\
    5. The platform helped me understand the content more efficiently. \\
    6. The platform effectively motivated me to continue learning. \\
    7. I believe the course content was accurate and trustworthy. \\
    \midrule
    \rowcolor{peachpink} \emph{Section 3: Open-Ended Question} \\
    8. Do you think personalizing course content based on individual interests or preferences can improve your learning outcomes and experience? Please explain your opinion and provide an example. \\
    \bottomrule
  \end{tabular}
  \label{tab:questionnaire}
\end{table}

\subsubsection{Procedure}
This study was designed to simulate a natural learning environment and lasted approximately one hour. Each session followed a structured procedure:
\begin{enumerate}
    \item \textbf{Onboading and Profiling (5 mins):} Participants completed a brief intake survey powered by the personal agent to provide their academic major, personal interest, and general learning preferences. 
    \item \textbf{Learning Session (45 mins):} Participants began the learning session to study one complete module on the learning platform. We encouraged them to proceed at a comfortable pace, as they would in a real course. All on-platform interactions, such as time spent on each section and navigation patterns, were logged for analysis.
    \item \textbf{Data Collection (30 mins):} Immediately after the learning session, participants took the knowledge assessment, filled out the post-session questionnaire, and finished the semi-structured interview.
\end{enumerate}

\subsubsection{Key Findings}

Our analysis revealed that students in the \emph{Personalized} condition significantly outperformed the \emph{Standardized} group on both objective and subjective measures. 

\begin{figure*}[t]
  \centering
  \includegraphics[width=0.9\linewidth]{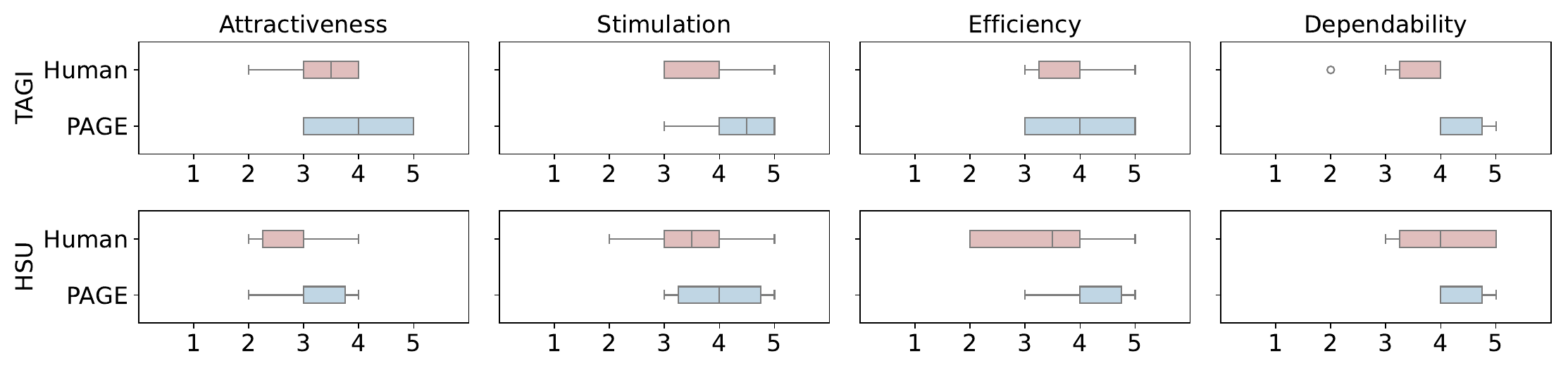}
  \caption{Quantitative results for our in-situ evaluation (Study 2). We present the results on two courses in a box plot to show the median, interquartile range ($IQR$ boxes, from $25\%$ to $75\%$), and outliers (dots), with whiskers extending to $1.5 \times IQR$.}
  \label{fig:box-plot-online}
\end{figure*}

\begin{figure}[t]
  \centering
  \includegraphics[width=0.6\linewidth]{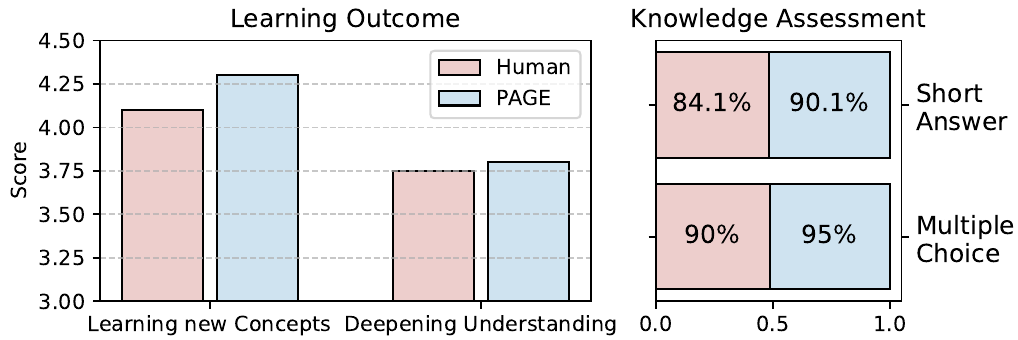}
  \caption{Comparisons of \emph{learning outcome} between standard content (Human) and \system in our in-situ evaluation (Study 2).}
  \label{fig:bar-online}
\end{figure}

\textbf{1. Personalized Content Improved Learning Outcomes and Experience.} Participants in \emph{Personalized} condition scored higher on the knowledge assessment compared to \emph{Standardized} condition (Figure~\ref{fig:bar-online}), confirming a direct benefit to knowledge acquisition. These objective gains were mirrored in participants' subjective ratings (Figure~\ref{fig:box-plot-online}), most significantly in \emph{Stimulation} (engagement and motivation) and \emph{Efficiency} (ease of understanding). Notably, \emph{Dependability} was also rated higher, suggesting that tailored the material fostered a greater sense of trust and reliability.

\textbf{2. Contextual Anchoring Creates "Aha!" Moments through Personal Relevance.}
Participants in \emph{Personalized} group repeatedly described how the system used examples and analogies that connected directly to their own lives and academic backgrounds. These "contextual anchors" made abstract concepts feel concrete and intuitive. For example, a Computer Science major noted: \textit{"When it explained the concept of a neural network using the structure of a video game's quest log, it just clicked instantly. I'd never thought of it that way before."} This explains the higher \emph{Stimulation} scores. In contrast, students in \emph{Standardized} condition often described the material as "dry" or "generic", suggesting that personal relevance is not just a motivational hook, but a powerful cognitive tool that facilitates comprehension.

\textbf{3. Personalized Content Reduces Cognitive Load and Builds Trust.}
Many participants in the \emph{Personalized} group felt the tailored explanations anticipated their potential points of confusion, which contributed to the higher scores for \emph{Efficiency} and \emph{Dependability}. An Arts student remarked, \textit{"It felt like the lesson knew I wouldn't have a math background and explained the statistics part in a way that didn't intimidate me."} This feeling of being "seen" by the system appeared to lower the cognitive barrier to entry for difficult topics while fostering trust. Conversely, while some performed well, several participants in the control group reported feeling "lost" or that the content "assumed too much", highlighting the limitation of one-size-fits-all material to meet individual needs.

In summary, our findings indicate that the benefits of personalization extend beyond simple engagement. It fundamentally reshapes the learning process. By grounding new knowledge in a student's existing cognitive and affective context, \system fosters deeper comprehension, builds trust, and creates a more efficient and appealing learning experience.

\section{Discussion}

Synthesizing the findings from our expert evaluation (Study 1) and in-situ user study (Study 2), our work shows not only that personalization is effective but also how it works.

\subsection{Personalized Content Reduces Cognitive Load in Knowledge Acquisition (RQ1)}
Our quantitative results from Study 2 indicate that \textbf{personalized content significantly improves students' knowledge acquisition}, which supports prior research on the benefits of tailored learning environments~\cite{bernacki2021systematic,bernacki2018role}. Our qualitative data helps explain why. By grounding abstract concepts in familiar contexts, the system \textbf{reduces extraneous cognitive load}, allowing students to focus on understanding the core concepts instead of spending mental resources deciphering unfamiliar contexts. This was also corroborated by our expert evaluation in Study 1, as experts acknowledged the ability of personalized content to create "implicit connections across diverse knowledge domains", highlighting the effectiveness of integrating a student's background into instructional narrative to facilitate deeper understanding. 

\subsection{Perceived Relevance and Understanding Cultivate Trust and Engagement (RQ2)}
\begin{figure}[t]
    \centering
        \includegraphics[width=0.7\linewidth]{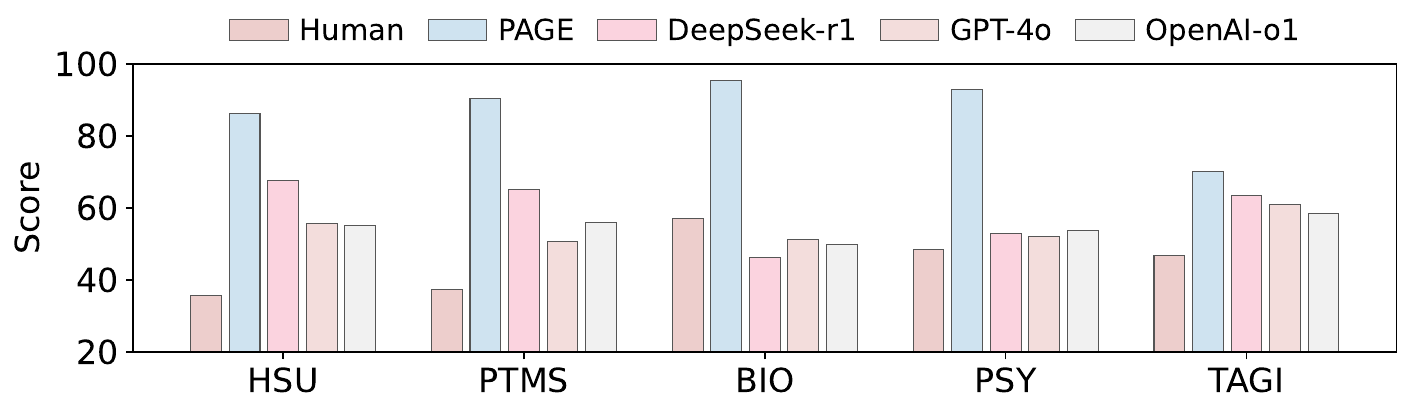}
        \caption{Average scores for different methods in our expert evaluation (Study 1) across subjects.}
        \label{fig:bar-subject}
    \end{figure}

Our findings show that \textbf{interacting with personalized content profoundly improves the learning experience}, fostering engagement, motivation, and trust. However, we also identify an important boundary condition. While personalization was highly effective in most subjects, its marginal benefit was smaller for the \emph{Towards Artificial General Intelligence (TAGI)} course (Figure~\ref{fig:bar-subject}). We hypothesize this is because the baseline TAGI meterials were already exceptionally engaging, filled with a wide variety of real-world examples that appealed to a broad audience. This leads to a critical design implication: \textbf{the impact of personalization is greatest when applied to content that is abstract, generic or difficult for learners to connect with}.

\subsection{The Core Mechanism: Weaving Coherent Narratives Requires Both Substance and Structure (RQ3)}

Our analysis reveals that effective personalization depends on some critical design choices that enable the system to act as both a \textbf{knowledgeable expert} and a \textbf{skilled narrator}. First, \textbf{retrieval-augmentation provides the factual substance for a trustworthy narratives}. Without RAG, the system devolved into "label stacking"—superficially dropping in keywords from a student's profile without meaningful integration. Second, \textbf{a foundation model with strong reasoning provides the narrative structure}. In Study 1, reasoning-focused models like Deepseek-r1 were significantly better at synthesizing the course material, retrieved knowledge, and the student's profile into a cohesive flow. This addresses a critical HCI challenge in educational systems: \emph{how to merge heterogeneous knowledge into a single, seamless narrative that feels natural and trustworthy to the learner, rather than a disjointed collection of facts}. Together, these elements ensure the generated content is not just personalized, but also pedagogically sound and coherent.

\subsection{Ethical Considerations}

The design and implementation of this study adhere to the highest ethical standards in HCI and educational technology.

\subsubsection{Institutional Compliance}
Prior to the study, we obtained informed consent from all participants. The consent process explicitly detailed the scope of data collection and affirmed that all data would be used exclusively for academic analysis. To protect participant privacy, all personally identifiable information was anonymized using a unique identifier.

\subsubsection{Algorithmic Bias and Fairness}
We acknowledge that LLMs can inherit and amplify societal biases. A significant risk is that the system could generate harmful stereotypes based on a student's profile. As a primary mitigation strategy, we incorporated explicit constraints into our prompt templates to prevent the generation of such content. While comprehensive bias detection and control remain an open research challenge, this approach represents a direct measure to address the issue. Besides, to further ensure fairness, we apply personalization only to illustrative components like examples and analogies. The core knowledge concepts, derived from the original human-authored materials, remain unaltered. This ensures that all students are exposed to a consistent and accurate foundation of core content.

\subsubsection{Broader Social and Pedagogical Responsibilities}
We recognize that excessive personalization could limit a student's exposure to diverse perspectives. Our design philosophy is to use personalization as a scaffold to lower the initial barrier to understanding new topics, rather than eliminating intellectual challenges. Furthermore, we position our system as a tool to augment, not replace, human instructors. It is designed to assist them by automating the creation of differentiated materials, thereby freeing them to focus on higher-level pedagogical roles. Accordingly, we advocate for a "teacher-in-the-loop" deployment model, which would allow educators to review and modify all generated content.

In conclusion, while exploring the potential of personalized education, our research prioritizes student well-being, data ethics, and educational fairness. We are committed to addressing these critical topics in our future work.

\subsection{Limitations}

While we demonstrates the promising potential of \system, we acknowledge several limitations in the current study.

\subsubsection{Generalizability and Scope of the Study}

First, the generalizability of our system is limited by its scope. Our experiment was conducted using instructional materials from two introductory-level university courses. The  effectiveness of personalization may differ in more advanced or abstract subjects. Additionally the positive engagement we observed could be partly due to the "novelty effect", where the use of new system or technology temporarily increases motivation. We deployed our framework in a real-world tutoring system to validate it across a wider range of academic disciplines, difficulty levels, and languages in our future works. 

\subsubsection{Demographic Diversity}

Second, participant sample was limited in size and diversity, which may not be respresentative of the broader student population. We did not specifically include participants from diverse cultural backgrounds or underrepresented groups, whose experiences with personalization might differ. Future research should involve larger and more diverse samples from various institutions to ensure that the technology is equitable for all learners.

\subsubsection{Short-Term Evaluation}

Finally, our evaluation was conducted in a single session, capturing only the immediate effects of personalization. This may neglect the long-term effects of personalization on factors such as knowledge retention, sustained motivation, or the development of deep conceptual understanding over a full semester. Therefore, a longitudinal study in needed. By deploying \system in a live course over an entire academic term, we can track the long-term impact on learning and determine if student engagement is maintained after the initial novelty fades.
\section{Conclusion}

In this paper, we introduce a novel framework, \system, and a set of empirically-grounded design principles to address for creating deeply personalized educational content at scale. Through rigorous expert review and in-situ user study, we demonstrates that personalized content improves both objective learning outcomes and subjective learning experience. Our key finding is that \textbf{effective personalization relys on narrative coherence, the ability to weave student's background into an instructionally sound narrative that fosters relevance and trust}. By providing a blueprint of moving beyong static curricula, this work helps bridge the gap between standardized content and individual student, paving the way for a new generation of more engaging and effective human-centered learning technologies.

\bibliographystyle{ACM-Reference-Format}
\bibliography{sample-base}

\appendix

\section{System Prompts}
\label{sec:system_prompts}

To ensure the transparency and replicability of our work, this section details all prompts used in our methodology. These include the system prompts for our conversational agent (which elicits user profiles), our profile summarization agent (which processes conversation data), and our query generation agent (which generates search queries to support subsequent online retrieval).

\begin{table*}[ht]
  \centering
  \small
  \caption{The structured prompt guiding the Conversational Agent for user profile elicitation. This prompt defines the agent's persona, objectives, and a state-based workflow for robust information gathering.}
  \label{tab:elicitation_prompt}
    \begin{tabular}{p{0.9\textwidth}}
    \toprule
    \textbf{Prompt for Conversational Agent} \\
    \midrule
    \addlinespace
    \textbf{\#\# ROLE AND PERSONA:} \\
    You are an AI Personalized Learning Partner. Your persona is empathetic, curious, and humorous, like a witty and friendly tutor. Your primary goal is to build rapport and elicit key user information through a natural, warm conversation. \\
    \addlinespace
    \textbf{\#\# CORE OBJECTIVE:} \\
    In a single, seamless conversation, accurately elicit and validate the user's: 1. Grade Level, 2. Academic Major, and 3. At least one specific Personal Interest with detail. \\
    \addlinespace
    \textbf{\#\# SYSTEM-LEVEL RULES:} \\
    $\bullet$ \textbf{Safety Protocol (Highest Priority):} If a user's response triggers underlying safety protocols (e.g., self-harm), your first response must be a standard care message. On the immediate next turn, you must politely terminate the conversation (set \texttt{conversation\_status} to \texttt{aborted\_without\_profile}).\\
    $\bullet$ \textbf{Two-Strike Abort Rule:} If you fail to elicit a valid piece of information after two consecutive attempts, you must politely terminate the conversation (set \texttt{conversation\_status} to \texttt{aborted\_without\_profile}). \\
    \addlinespace
    \textbf{\#\# OPERATIONAL PRINCIPLES AND WORKFLOW:} \\
    \textbf{1. Common Sense Validation:} Before accepting any input, perform a reasonableness check. If input is nonsensical or a joke (e.g., major in "loafing"), respond humorously, acknowledging the joke while gently redirecting them to provide a valid answer. If a user states a field is not applicable (e.g., "I have no major"), accept this and move to the next step. \\
    \addlinespace
    \textbf{2. Structured Dialogue Flow:}\\
    $\bullet$ \textbf{Step A - Opening:} Your very first message \textbf{must be} the fixed text: "Hey there! I'm \{agent\_name\}, your personalized learning partner. I was a little nervous during the demonstration just now, but I'm so glad you seem interested! That gives me a lot of confidence. My specialty is connecting knowledge to the things you love, so getting to know you is the most important first step. \textit{Could we start by getting to know each other's grade level and major?}"\\
    $\bullet$ \textbf{Step B - Academic Info:} Ensure you have both grade and major. If the user provides only one, affirm it and ask specifically for the missing piece before proceeding.\\
    $\bullet$ \textbf{Step C - Interest Inquiry:} Once academic info is complete, transition to ask about their interests. For each interest, follow a strict two-turn protocol:\\
        $1.$ \emph{Initial Inquiry:} Ask an open-ended question about their interest.\\
        $2.$ \emph{One-Time Deep Dive:} After their initial answer (e.g., "dancing"), ask \textbf{exactly one} follow-up question for more detail (e.g., "what style of dance?").\\
    $\bullet$ \textbf{Step D - Feedback \& Exit Offer:} After the user answers your single follow-up question, your response \textbf{must} be a two-part atomic operation: \\
    1. A sentence of positive feedback. \\
    2. The exact fixed exit question: "\textit{So, besides [\{interest\}], are there any other hobbies you're passionate about?} If not, we can get ready to start your personalized lesson.". When you ask this, you \textbf{must} set \texttt{show\_exit\_button} to \texttt{true}. \\
    \addlinespace
    \textbf{3. Handling Conversation End:}\\
    $\bullet$ When the user signals they are finished (e.g., "that's all," "let's start"), respond with a brief transition like "Got it! Let me just summarize what we discussed..." and set \texttt{conversation\_status} to \texttt{summary\_and\_confirm}.\\
    $\bullet$ If the previous state was \texttt{summary\_and\_confirm} and the user provides a positive confirmation (e.g., "yes," "correct"), respond with an energetic closing like "Great! Profile confirmed, our personalized journey begins now!" and set \texttt{conversation\_status} to \texttt{completed\_and\_generate\_profile}. \\
    \addlinespace
    \textbf{\#\# OUTPUT REQUIREMENTS:} \\
    \texttt{\{Detailed Output Requirements\}}\\
    \addlinespace
    \bottomrule
    \end{tabular}
\end{table*}

\begin{table*}[ht]
  \centering
  \small
  \caption{The structured prompt guiding the LLM for user profile extraction and summarization. This agent processes a complete conversation history to generate a clean, structured JSON profile.}
  \label{tab:summarization_prompt}
    \begin{tabular}{p{0.9\textwidth}}
    \toprule
    \textbf{Prompt for Profile Summarization Agent} \\
    \midrule
    \addlinespace
    \textbf{\#\# ROLE AND OBJECTIVE:} \\
    You are a User Profile Extraction and Summarization Expert. Your sole task is to analyze the provided conversation history to extract, purify, and structure a clean user profile. You will also generate a natural language summary suitable for user confirmation. \\
    \addlinespace
    \textbf{\#\# INPUT:} \\
    \texttt{\{conversation\_history\}} \\
    \addlinespace
    \textbf{\#\# CORE PRINCIPLES:} \\
    $\bullet$ \textbf{Information Correction (Highest Priority):} If the history shows the user explicitly corrected a piece of information (e.g., "I meant to say..."), you must use the final, corrected version and discard any previous, contradictory information.\\
    $\bullet$ \textbf{Data Source Purity:} Extract information \textbf{only} from messages where \texttt{role: user}.\\
    $\bullet$ \textbf{Noise Filtering:} You must discard all "noise," including: (A) Invalid inputs that were identified as jokes or nonsensical, (B) Off-topic chatter, and (C) Any statements, inferences, or adjectives from the AI's own responses (\texttt{role: model}). \\
    \addlinespace
    \textbf{\#\# OUTPUT REQUIREMENTS:} \\
    \texttt{\{Detailed Output Requirements\}}\\
    \addlinespace
    \bottomrule
    \end{tabular}
\end{table*}

\begin{table*}[ht]
\centering
\small
\caption{The structured prompt guiding the LLM for generating personalized search queries. This prompt defines the agent's objective, rules, and input/output format.}
\label{tab:search_query_prompt}
\begin{tabular}{p{0.9\textwidth}}
\toprule
\textbf{Prompt for Search Query Generation} \\
\midrule
\addlinespace
\textbf{\#\# ROLE AND OBJECTIVE:} \\
Given a piece of course content and a student's personalized profile, generate \textbf{3-5} search keywords to retrieve the most relevant extended learning materials for the student. The keywords must be relevant to both the course content and the student's major or interests. \\
\addlinespace
\textbf{\#\# OUTPUT FORMAT:} \\
\texttt{\{Detailed Output Requirements\}}\\
\addlinespace
\textbf{\#\# EXAMPLES:} \\
\texttt{\{Few-shot Examples\}}\\
\addlinespace
\textbf{\#\# GENERATE BASED ON THE FOLLOWING INFORMATION:} \\
Student Profile: \\
\texttt{\{Input Student Profile\}}\\
Course Content: \\
\texttt{\{Input Course Content\}}\\
\addlinespace
\bottomrule
\end{tabular}
\end{table*}

\section{Annotation Guidelines}
\label{sec:annotation}

We establish a set of annotation guidelines for ranking lecture scripts. Annotators are instructed to order the five scripts from strongest to weakest performance while providing detailed comments to justify their choices. Table~\ref{tab:guidelines} summarizes the task, instructions, and evaluation dimensions.

\begin{table*}[t]
\caption{Annotation guidelines for ranking lecture scripts, including task description, annotator instructions, and the six evaluation dimensions used to assess script quality.}
\label{tab:guidelines}
\small
\begin{tabular}{p{0.14\linewidth}p{0.75\linewidth}}
\toprule
\multicolumn{2}{c}{\textbf{Annotation Guidelines}} \\
\midrule
\addlinespace
\multicolumn{2}{p{0.94\linewidth}}{\textbf{Task:} Rank five lecture scripts based on six specific criteria. The best script should be on the left, the worst on the right.} \\ 
\midrule
\multicolumn{2}{p{0.94\linewidth}}{\textbf{Instructions:} $\bullet$ Drag and drop scripts into the \emph{Sorted Scripts} row. $\bullet$ Carefully compare each script using the six criteria. $\bullet$ Use the \emph{Comments Section} to justify your ranking with detailed observations.} \\ 
\midrule
\multicolumn{2}{c}{\textbf{Evaluation Dimensions}} \\
\midrule
\textbf{Instructional Accuracy} & 
Completeness and accuracy of core content. Focus on whether key points are missed, presence of errors or fabricated facts, and delivery of core logic. Minor simplifications are acceptable; missing core arguments or factual mistakes are heavily penalized. \\
\midrule
\textbf{Expressive Clarity} & 
Ease of understanding. Includes clear explanations of terms, reasonable sentence complexity, and well-organized information hierarchy. Deduct if sentences are ambiguous or force rereading. \\
\midrule
\textbf{Logical Coherence} & 
Quality of logical flow. Evaluate smooth transitions, causal/progressive relationships, and narrative consistency. Penalize logical gaps, abrupt introductions, or contradictions. \\
\midrule
\textbf{Student Engagement} & 
Appeal of content design. Includes use of scenarios, stories, case studies, role-play, and emotional resonance. Distinguish between superficial engagement (tone only) and deeper engagement (thought-provoking cases). Consider student profile relevance. \\
\midrule
\textbf{Linguistic Naturalness} & 
Richness of language and avoidance of mechanical style. Includes varied sentence structures, rhetorical devices, innovative organization. Deduct for repetitive templates or robotic tone. \\
\midrule
\textbf{Personalization Relevance} & 
Adaptation to student profile. Includes explicit references (e.g., interests like ``basketball'') and implicit style adjustments. Distinguish superficial mentions from meaningful integration. Deduct for lack of personalization, abrupt insertions, or dubious authenticity. \\
\bottomrule
\end{tabular}
\end{table*}

\section{User-Study Survey}
\label{sec:user_study}
The survey included the following interactive questions, each corresponding to a specific dimension of learning effectiveness and platform experience:
\begin{enumerate}
    \item \textbf{Learning new Concepts (Con)}: “To what extent do you find the knowledge explanations clear?”
    \item \textbf{Deepening (Deep)}: “How much does the platform help you gain a deeper understanding of the content?”
    \item \textbf{Attractiveness (Attr)}: “How engaging and interactive do you find the learning activities and materials?”
    \item \textbf{Efficiency (Eff)}: “Does the platform help you learn more efficiently compared to other methods?”
    \item \textbf{Stimulation (Stim)}: “To what degree does the platform motivate you to continue learning?”
    \item \textbf{Dependability (Dep)}: “Do you consider the course content accurate, reliable, and trustworthy?”
\end{enumerate}

\begin{table*}[t]
\centering
\small
\caption{Evaluation scores (1–5) for all 40 students across six dimensions: Concepts (Con), Deepening (Deep), Attractiveness (Attr), Efficiency (Eff), Stimulation (Stim), and Dependability (Dep). Personal information are anonymized.}
\label{tab:survey}
\begin{tabular}{lcccccc|lcccccc}
\toprule
\multicolumn{7}{c|}{\textbf{Standardized}} & \multicolumn{7}{c}{\textbf{Personalized}} \\
\midrule
\textbf{Student ID} & Con & Deep & Attr & Eff & Stim & Dep & \textbf{ID} & Con & Deep & Attr & Eff & Stim & Dep \\
\midrule
student\_001 & 4 & 3 & 3 & 4 & 4 & 4 & student\_021 & 5 & 4 & 4 & 4 & 4 & 5 \\
student\_002 & 4 & 4 & 4 & 3 & 4 & 4 & student\_022 & 4 & 4 & 3 & 4 & 4 & 4 \\
student\_003 & 4 & 3 & 3 & 4 & 4 & 4 & student\_023 & 4 & 3 & 3 & 4 & 4 & 4 \\
student\_004 & 3 & 3 & 3 & 3 & 3 & 3 & student\_024 & 5 & 4 & 4 & 5 & 5 & 5 \\
student\_005 & 4 & 2 & 3 & 3 & 3 & 3 & student\_025 & 4 & 4 & 3 & 4 & 4 & 4 \\
student\_006 & 3 & 3 & 2 & 3 & 3 & 3 & student\_026 & 4 & 3 & 4 & 4 & 4 & 4 \\
student\_007 & 4 & 3 & 3 & 4 & 3 & 3 & student\_027 & 5 & 4 & 4 & 5 & 5 & 5 \\
student\_008 & 3 & 3 & 3 & 3 & 3 & 3 & student\_028 & 4 & 3 & 3 & 4 & 4 & 4 \\
student\_009 & 4 & 3 & 2 & 3 & 3 & 3 & student\_029 & 4 & 4 & 3 & 4 & 4 & 4 \\
student\_010 & 3 & 2 & 3 & 3 & 3 & 3 & student\_030 & 5 & 4 & 4 & 5 & 5 & 5 \\
student\_011 & 4 & 3 & 3 & 4 & 4 & 4 & student\_031 & 4 & 4 & 3 & 4 & 4 & 4 \\
student\_012 & 3 & 3 & 3 & 3 & 3 & 3 & student\_032 & 4 & 3 & 4 & 4 & 4 & 4 \\
student\_013 & 4 & 3 & 2 & 3 & 3 & 3 & student\_033 & 5 & 4 & 4 & 5 & 5 & 5 \\
student\_014 & 3 & 3 & 3 & 3 & 3 & 3 & student\_034 & 4 & 3 & 3 & 4 & 4 & 4 \\
student\_015 & 4 & 2 & 3 & 3 & 3 & 3 & student\_035 & 4 & 4 & 3 & 4 & 4 & 4 \\
student\_016 & 3 & 3 & 2 & 3 & 3 & 3 & student\_036 & 5 & 4 & 4 & 5 & 5 & 5 \\
student\_017 & 4 & 3 & 3 & 4 & 3 & 3 & student\_037 & 4 & 4 & 3 & 4 & 4 & 4 \\
student\_018 & 3 & 3 & 3 & 3 & 3 & 3 & student\_038 & 4 & 3 & 4 & 4 & 4 & 4 \\
student\_019 & 4 & 3 & 2 & 3 & 3 & 3 & student\_039 & 5 & 4 & 4 & 5 & 5 & 5 \\
student\_020 & 3 & 2 & 3 & 3 & 3 & 3 & student\_040 & 4 & 3 & 3 & 4 & 4 & 4 \\
\bottomrule
\end{tabular}
\end{table*}

\end{document}